\documentclass[prl,twocolumn,preprintnumbers,amsmath,amssymb]{revtex4}

\usepackage{graphicx}

\begin{document}
\title{ \bf  Quantum phase transitions from Solids to Supersolids in bi-partite lattices }
\author{ \bf  Jinwu Ye  }
\affiliation{ Department of Physics, The Pennsylvania State
University, University Park, PA, 16802 }
\date{\today}

\begin{abstract}
    We studied some phases and phase transitions in an
    extended boson Hubbard model
    slightly away from half filling on bipartite lattices such as honeycomb and square lattice.
    We find that in the insulating side, different kinds of supersolids are generic
    stable states slightly away from half filling. We propose a new kind of supersolid: valence bond
    supersolid. We show that the quantum phase transitions from solids to
    supersolids driven by a chemical potential are in the
    same universality class as that from a Mott insulator to a superfluid,
    therefore have exact exponents $ z=2, \nu=1/2, \eta=0 $ ( with logarithmic corrections ).
    Comparisons with previous quantum Monte-Carlo (QMC) simulations on
    some microscopic models on a square lattice are made.
    Implications on possible future QMC simulations  are given.
\end{abstract}

\maketitle

     {\sl 1. Introduction: } A supersolid is a state with both
     superfluid and solid order.
     Recently, by using the torsional oscillator measurement, a PSU's group lead by Chan
     observed a marked $ 1 \sim 2 \% $  Non-Classical Rotational Inertial  ( NCRI)
     even in bulk solid $^{4} He $ at $ \sim 0.2 K $  \cite{chan}.
     The NCRI is a low temperature reduction in the rotational moment of inertia
     due to the superfluid component of the state \cite{leg}.
     If this experimental observation indicates the existence of $^{4} He $ supersolid
     remains controversial \cite{qgl,qgll}. However,
     it was established by spin wave expansion \cite{gan}
     and quantum Monte-carlo (QMC) \cite{hard,add,soft} simulations that
     a supersolid state could exist in an extended boson Hubbard model (EBHM) with suitable lattice structures,
     filling factors, interaction ranges and strengths.
     But so far, the universality classes of the quantum phase transitions from  solids to supersolids have never
     been studied. In this letter, by using the dual vortex method developed in \cite{pq1},
     we investigate some
     phases, especially supersolids and quantum phase transitions in an extended boson Hubbard model (EBHM) on bipartite lattices such as
     honeycome and square lattice near half filling.
     Although the SS in lattice models is different
     from that in a continuous systems,
     the results achieved in this paper on lattice supersolids may still shed some lights on the possible
     microscopic mechanism and phenomenological Ginsburg-Landau
     theory of the possible $ ^{4}He $ supersolids \cite{qgll}.
     The EBHM in honeycomb and square lattices could be realized in
     ultracold atoms loaded on optical lattices.
     So the results achieved in this paper may have direct impacts on the atomic experiments.

     The EBHM with various kinds of interactions,
     on all kinds of lattices and at different filling factors is described by the following
     Hamiltonian \cite{boson}:
\begin{eqnarray}
   H  & = & -t \sum_{ < ij > } ( b^{\dagger}_{i} b_{j} + h.c. )
          - \mu \sum_{i} n_{i} + \frac{U}{2} \sum_{i} n_{i} ( n_{i} -1 )
                                  \nonumber   \\
      & + & V_{1} \sum_{ <ij> } n_{i} n_{j}  + V_{2} \sum_{ <<ik>> } n_{i} n_{k} + \cdots
\label{boson}
\end{eqnarray}
    where $ n_{i} = b^{\dagger}_{i} b_{i} $ is the boson density, $ t $ is the nearest neighbor hopping amplitude.
    $ U, V_{1}, V_{2} $ are onsite, nearest neighbor (nn) and next nearest neighbor (nnn) interactions respectively,
    the $ \cdots $ may include further neighbor interactions and possible ring-exchange interactions.
     A supersolid  is defined as the
     simultaneous orderings of ferromagnet in the $ XY $ component and CDW in the $ Z $ component.
     Honeycomb lattice is not a Bravais lattice, so may show some different properties
     than those in a square lattice.

     {\sl 2. The dual vortex method. }
     The Eqn.\ref{boson} with only the onsite interaction was first studied in Ref.\cite{boson}.
     The effects of long range Coulomb interactions on the transition was studied in \cite{yeboson}.
     Very recently, the most general cases Eqn.\ref{boson} in a square lattice
     at generic commensurate filling factors $ f=p/q $ ( $ p, q $ are relative prime numbers ) were
     systematically studied in \cite{pq1}.
    After performing the charge-vortex duality transformation, the authors in \cite{pq1}
    obtained a dual theory of Eqn.\ref{boson}
    in term of the interacting vortices $ \psi_{l} $ hopping on the dual lattice subject to a fluctuating
    {\em dual} " magnetic field".
    The average strength of the dual " magnetic field "  through a dual plaquette is equal to the boson density $ f=p/q $.
    This is similar to the Hofstadter problem
    of electrons moving in a crystal lattice in the presence of a magnetic field.
    The magnetic space group (MSG) in the presence of this dual magnetic field
    dictates that there are at least $ q $-fold degenerate minima in the mean field energy spectrum.
    The $ q $ minima can be labeled as $ \psi_{l}, l=0,1,\cdots, q-1 $ which forms a $ q $ dimensional
    representation of the MSG.  In the continuum limit, the final effective theory describing
    the superconductor to the insulator transition in terms of these $ q $ order parameters should be
    invariant under this MSG. In this letter, we will extend the dual vortex method to
    study the EBHM Eqn.\ref{boson} in honeycomb lattice at and {\em slightly away } from $ q=2 $ (Fig.1a)

        The dual approach is a MSG symmetry-based approach which can be used to classify some phases
        and phase transitions. But the question if a particular phase will appear or
        not as a ground state can not be addressed in this
        approach, because it depends on the specific values of all the
        possible parameters in the EBHM in Eqn.\ref{boson}.
        So a microscopic approach such as Quantum Monte-Carlo (QMC)
        may be needed to compare with the dual field theoretical approach.
        The dual approach can guide the QMC to search for particular phases
        and phase transitions in a specific model. Finite size
        scalings in QMC can be used to confirm the universality
        class discovered by the dual approach.



   {\sl 3. The effective action and order parameters in the dual vortex picture.}  The dual lattice of the honeycomb lattice is a triangular lattice.
   Two basis vectors of a primitive unit cell  of the triangular lattice can be chosen as
   $ \vec{a}_{1}=  \hat{x}, \vec{a}_{2}= - \frac{1}{2} \hat{x} + \frac{ \sqrt{3} }{ 2 } \hat{y},
    \vec{a}_{d}= \vec{a}_{1}+ \vec{a}_{2} $ as shown in Fig.1a. The reciprocal lattice of a triangular
   is also a triangular lattice and spanned by two basis vectors
   $ \vec{k}= k_{1} \vec{b}_{1} + k_{2} \vec{b}_{2} $ with $ \vec{b}_{1}= \hat{x} + \frac{ \hat{y} }{ \sqrt{3} },
     \vec{b}_{2}= \frac{2}{\sqrt{3}} \hat{y} $ satisfying $ \vec{b}_{i} \cdot \vec{a}_{j} = \delta_{ij}  $.
    The point group of a triangular lattice is $ C_{6v} \sim D_6 $ which contains $ 12 $ elements. The two generators
    can be chosen as $ C_{6} = R_{\pi/3},  I_{1} $. The space group also includes the two translation operators
    along $ \vec{a}_{1} $ and $ \vec{a}_{2} $ directions $ T_{1} $ and $ T_{2} $.
    The 3 translation operators  $ T_{1}, T_{2}, T_{d} $, the rotation operator $ R_{\pi/3} $,
    the 3 reflection operators $ I_{1}, I_{2}, I_{d} $, the two rotation operators around the direct
    lattice points $ A $ and $ B $: $   R^{A}_{2\pi/3}, R^{B}_{2\pi/3} $ of the MSG
    are worked out in \cite{un}.  It can be shown that they all commute with $ {\cal H}_{v} $.
    However, they do not commute with each other, for example,
    $ T_{1} T_{2}= \omega T_{d}, T_{1} T_{2}= \omega^{2} T_{2}T_{1} $ where $ \omega= e^{i 2 \pi f }$.

    In the following, we focus on $ q=2 $ case \cite{un} where there is only {\em one } dual vortex band
   $ E( \vec{k} ) = - 2 t ( \cos k_{1} + \cos k_{2} - \cos ( k_{1} + k_{2} ) ) $. Obviously,
   $ E( k_1, k_2 ) = E( - k_1, -k_2 ) = E( k_2, k_1 ) $.  There are {\em two }  minima at $ \vec{k}_{\pm}=
   \pm ( \pi/3,\pi/3 ) $. Let's label the two eigenmodes at the two minima as
   $ \psi_{a/b} $.  How the two fields transform under the MSG was derived in
   \cite{un}.

    Moving {\em slightly} away from half filling $ f=1/2 $ corresponds to adding
    a small {\em mean} dual magnetic field $ H \sim  \delta f= f-1/2 $ in the action.
    It can be shown that {\em inside the SF phase }, the most general
    action invariant under all the MSG transformations  upto quartic terms is \cite{un}:
\begin{eqnarray}
    {\cal L}_{SF} & = & \sum_{\alpha=a/b} | (  \partial_{\mu} - i A_{\mu} ) \psi_{\alpha} |^{2} + r | \psi_{\alpha} |^{2}
    +  \frac{1}{4 e^{2} } ( \epsilon_{\mu \nu \lambda} \partial_{\nu} A_{\lambda}
    - 2 \pi \delta f \delta_{\mu \tau})^{2}   \nonumber  \\
       & +  & \gamma_{0} ( | \psi_{a} |^{2} + |\psi_{b} |^{2} )^{2} -
                       \gamma_{1} ( | \psi_{a} |^{2} - |\psi_{b} |^{2} )^{2} + \cdots
\label{away}
\end{eqnarray}
     where $ A_{\mu} $ is a non-compact  $ U(1) $ gauge field, e is a dimensionless coupling constant
     depending on $ t, U, V_{1}, V_{2} \cdots $ in Eqn.\ref{boson}. Upto
     the quartic level, with correspondingly defined $ \psi_{a/b} $ in a square lattice,
     Eqn.\ref{away} is the same as that in the square lattice derived in \cite{pq1}.

     Because the duality transformation is a non-local
     transformation, the relations between the phenomenological
     parameters in Eqn.\ref{away} and the microscopic parameters in
     Eqn.\ref{boson} are highly non-local and not known.
     Fortunately, we are still able to classify some
     phases and phase transitions and make some very sharp predictions from Eqn.\ref{away} without knowing
     these relations.
     If $ r > 0 $, the system is in the superfluid state $  < \psi_{l} > =0 $ for every $ l=a/b $.
     If $ r < 0 $, the system is in the insulating state $ < \psi_{l} > \neq 0 $ for
     at least one $ l $.
    In the insulating or supersolid states, there must exist some kinds of charge density wave (CDW)
    or valence bond solid (VBS) orders which may be stabilized by longer range interactions or possible
    ring exchange interactions in Eqn.\ref{boson}. Up to an unknown prefactor
    \cite{un}, we can identify the boson ( or vacancy ) densities on sites A and B and
    the boson kinetic energy on the link between A and B
    which are the order parameters for the CDW and VBS respectively as
    $ \rho_{A}  =   \psi^{\dagger}_{a} \psi_{a},~ \rho_{B}=  \psi^{\dagger}_{b} \psi_{b} $
    and $ K_{AB}  =   e^{i \vec{Q} \cdot \vec{x} } \psi^{\dagger}_{a}
    \psi_{b} + e^{-i \vec{Q} \cdot \vec{x} }  \psi^{\dagger}_{b} \psi_{a} $
    where $ \vec{Q}= 2 \pi/3 (1,1) $ and  $ \vec{x} $ stands for dual lattice points {\em only}.
    We assume $ r < 0 $ in Eqn.\ref{away}, so the system is in the insulating state.
    In the following, we discuss the Ising limit first, then the easy-plane limit.

 {\sl 4. Phase diagram (a) Ising limit.}
   If $ \gamma_{1} > 0 $, the system is in the Ising limit, the mean field
   solution is $ \psi_{a} =1, \psi_{b}=0 $. The system is in the CDW order
   which could take checkboard $ (\pi,\pi) $ order \cite{equal}.
   Eqn.\ref{away} is an expansion around the  uniform saddle point $
   < \nabla \times \vec{A} > = f = 1/2 $ which holds in the SF and the
   VBS ( to be discussed in section 5 ). In the CDW state, a
   different saddle point where  $ <\nabla \times \vec{A}^{a}> = 1-\alpha
   $ for sublattice $ A $ and $ < \nabla \times \vec{A}^{b}> = \alpha
   $ for sublattice $ B $ should be used. So the transition from the SF to the CDW is a {\em strong} first
   order transition. It can be shown \cite{un} that there
   is only one vortex minimum $ \psi_{b} $ in such a staggered dual magnetic
   field with $ \alpha < 1/2 $, the effective action inside the CDW state is:
\begin{eqnarray}
  {\cal L}_{CDW} & = & | (  \partial_{\mu} - i A^{b}_{\mu} ) \psi_{b} |^{2} + r |
  \psi_{b}|^{2} +   u  | \psi_{b} |^{4} + \cdots     \nonumber  \\
     & + & \frac{1}{ 4 e^{2} } ( \epsilon_{\mu \nu \lambda} \partial_{\nu} A^{b}_{\lambda}
    - 2 \pi \delta f \delta_{\mu \tau})^{2}
\label{is}
\end{eqnarray}
   where the vortices in the phase winding of $ \psi_{b} $  should be interpreted as the
   the boson number \cite{direct}. The gauge field $ \vec{A}^{a} $
   is always massive.

   Eqn.\ref{is} has the structure identical to the conventional $ q=1 $ component
   Ginzburg-Landau model for a " superconductor "  in a "magnetic"
   field. By a duality transformation back to the boson $ \Psi $  picture,
   it can be shown that the transition driven by $ \delta f $ is
   in the same universality class of Mott to superfluid transition
   which has the exact exponents  $ z=2, \nu=1/2, \eta=0 $ with logarithmic corrections first discussed in \cite{boson,subir}.
   This fact was used in \cite{huse} to show that
   for type II superconductors, the gauge field fluctuations will render the vortex fluid phase
   intruding at $ H_{c1} $ between the Messiner and the mixed phase Fig2a.  For parameters appropriate to the cuprate
   superconductors, this intrusion occurs over too narrow an interval of $ H $ to be observed in experiments.
   In the present boson problem with the nearest neighbor interaction $
   V_{1} > 0 $ in Eqn.\ref{boson} which stabilizes
   the $ (\pi,\pi) $ CDW state at $ f=1/2$, this corresponds to a CDW  supersolid (CDW-SS) state intruding between
   the commensurate CDW state at $ f=1/2 $ and
   the in-commensurate CDW state at $ 1/2 + \delta f $ which could be stabilized by further neighbor
   interactions in Eqn.\ref{boson} as shown in Fig.2a.
   The first transition is in the $ z=2, \nu=1/2, \eta=0 $ universality
   class, while the second is a 1st order transition.
   We expect the intruding window at our $ q=2 $ system is much {\em wider } than that of the $ q=1 $ system.
   In the CDW-SS state, $ < \psi_{b} > =0 $, there is the
   gapless superfluid mode represented by the dual gauge field $
   A^{b}_{\lambda} $ in Eqn.\ref{is}, there is also the same $ ( \pi, \pi ) $ diagonal order
   as the C-CDW where both $ \vec{A}^{a} $ and $ \vec{A}^{b} $ are massive.
   For example, near the C-CDW to the CDW-SS transition, the superfluid
   density  should scale as $ \rho_{s} \sim |\rho-1/2|^{(d+z-2)\nu }=|
   \rho-1/2|= \delta f $ with logarithmic corrections.
   It is known the SF is stable against changing the chemical
   potential ( or adding bosons ) in Fig.1b. There
   must be a transition from the CDW-SS to the SF inside the window driven by
   the quantum fluctuation $ r $ in the Fig.1b. The universality
   class of this transition is likely to be first order and will be investigated in a future
   publication \cite{un}.

\begin{figure}
\includegraphics[width=8cm]{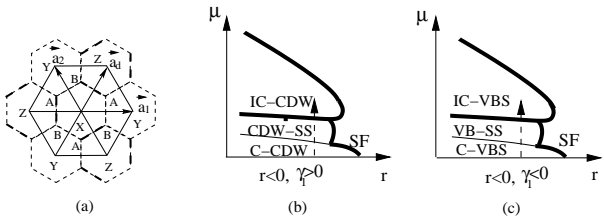}
\caption{ (a) Bosons at filling factor $ f $ are hopping on a
honeycomb lattice ( dashed line ) which has two sublattices $ A $
and $ B $. Its
    dual lattice is a triangular lattice ( solid line ) which has three sublattices $ X, Y, Z
    $. In the easy-plane limit,
    one of the three VBS states \cite{vbs} is shown in the figure.
    The thick dashed bond is twice as strong as that of the thin
    dashed bond \cite{un}. The other two VBS can be obtained by $ R^{A}_{ 2 \pi /3} $ or
    $ R^{B}_{ 2 \pi /3} $.
   (b) and (c) are the  zero temperature phase diagrams of
   the chemical potential $ \mu $ versus $ r $ in Eqn.\ref{away} in the honeycomb lattice.
   (b) The Ising limit $ \gamma_{1} > 0 $. There is a CDW  supersolid
   (CDW-SS) state intruding between the commensurate
   CDW ( C-CDW ) state at $ f=1/2 $ and the in-commensurate CDW (IC-CDW) state at $ 1/2 + \delta f $.
    The CDW-SS
    {\em has the same lattice symmetry breaking as the C-CDW }.
   (c) The Easy-Plane limit $ \gamma_{1} < 0 $. There is a Valence Bond Supersolid (VB-SS) state intruding between the commensurate
   VBS (C-VBS) state at $ f=1/2 $ and the in-commensurate VBS ( IC-VBS ) state at $ 1/2 +
   \delta f $. The VB-SS  {\em has the same lattice
   symmetry breaking as the C-VBS }. The thin ( thick ) line is the 2nd ( 1st ) order transition.
   The 1st order transition in the Ising ( Easy-plane ) limit is weakly ( strongly ) one. }
\label{fig1}
\end{figure}


 {\sl 5. Phase diagram (b) Easy-plane limit.} If $ \gamma_{1} < 0 $, the system is in the easy-plane limit.
   This limit could be reached by possible ring exchange
   interactions \cite{sand} in Eqn.\ref{boson}. At $ q=2 $, the mean field
   solution is $ \psi_{a}=e^{i \theta_{a} }, \psi_{b}= e^{i \theta_{b}} $.
   $ \rho_{A}= \rho_{B} = 1 $, so the two sublattices remain equivalent.
   The uniform saddle point $ < \nabla \times \vec{A} > = f = 1/2 $ holds in both the SF and the
   VBS, so the transition from the SF to the VBS is a {\em weak } first order transition.
   The system has a VBS order, the kinetic energy $ K_{AB} =  \cos( \vec{ Q } \cdot \vec{x} +
   \theta_{-} ) $ where $ \theta_{-} = \theta_{a} - \theta_{b} $.  Upto the quartic order,
   the relative phase $ \theta_{-} $ is undetermined. Higher order terms are needed
   to determine the relative phase.  It was shown in \cite{un} that there are only 3 sixth order invariants:
    $  C_{1}  =  | \psi_{a} |^{6} + |\psi_{b} |^{6},
       C_{2}  =  ( | \psi_{a} |^{2} + |\psi_{b} |^{2} ) | \psi_{a} |^{2} |\psi_{b} |^{2},
       C_{3}  =   ( \psi^{*}_{a} \psi_{b} )^{3}  + ( \psi^{*}_{b} \psi_{a} )^{3} = \lambda \cos 3 \theta_{-} $.
   Obviously, only the last term $ C_{3} $  can fix the relative phase. It is easy to show that both signs
   of $ \lambda $ are {\em equivalent}, in sharp contrast to the square lattice where the two signs of
   $  C_{sq}= \lambda \cos 4 \theta $ lead to either Columnar dimer or plaquette pattern.
   One VBS with $ \lambda > 0, \theta_{-} = \pi $ was shown in Fig.1a. The other two
   VBS states can be obtained by $ R^{A/B}_{ 2 \pi/3} $.  In contrast
   to the square lattice where $ C_{sq} $ is irrelevant near the QCP, $
   C_{3} $ may be relevant, so the transition from the SF to the VBS
   in Fig.1b could be a 1st order transition.
    Slightly away from the half-filling, Eqn.\ref{away} becomes:
\begin{eqnarray}
  {\cal L}_{VBS}  &  =  & ( \frac{1}{2} \partial_{\mu} \theta_{+} - A_{\mu} )^{2}
      + \frac{1}{4 e^{2}} ( \epsilon_{\mu \nu \lambda} \partial_{\nu} A_{\lambda}
      - 2 \pi \delta f \delta_{\mu \tau})^{2} + \cdots    \nonumber  \\
       & + &  ( \frac{1}{2} \partial_{\mu} \theta_{-} )^{2} + 2 \lambda \cos 3 \theta_{-}
\label{ep1}
\end{eqnarray}
   where $ \theta_{\pm} = \theta_{a} \pm \theta_{b} $.

   Obviously, the $ \theta_{-} $ sector is massive ( namely, $ \theta_{a} $ and $ \theta_{b} $ are {\em locked} together )
   and can be integrated out. Assuming $ \lambda > 0 $, then $ \theta_{-} = \pi $.
   Setting $ \psi_{+}=e^{i \theta_{+} } \sim \psi_{a} \sim -\psi_{b}$ in Eqn.\ref{away} leads
   to Eqn.\ref{is} with $ u = 2 \gamma_{0} $, so
   the discussions on Ising limit case following Eqn.\ref{is} also
   apply. In the present boson problem with possible ring exchange interactions in Eqn.\ref{boson} which
   stabilizes the VBS state at $ f=1/2$, this corresponds to a VBS supersolid (VB-SS) state intruding between the commensurate
   VBS ( C-VBS) state at $ f=1/2 $ and the in-commensurate VBS (IC-VBS) state at $ 1/2 +
   \delta f $ as shown in Fig.2b.  In this IC-VBS state, $ \delta f $
   valence bonds shown in Fig.1a is slightly stronger than the
   others. In the VB-SS state, $ < \psi_{a} > = < \psi_{b} > =0 $,
   but $ < \psi^{\dagger}_{a} \psi_{a} > = < \psi^{\dagger}_{b} \psi_{b} > = - < \psi^{\dagger}_{a} \psi_{b} > \neq 0 $,
   so there is a VBS order $ K_{AB} =  \cos( \vec{ Q } \cdot \vec{x} + \theta_{-} )
   $ which is the same as the C-VBS, the superfluid density $ \rho_{s} \sim \delta f $.
   Again, the first transition is in the  $ z=2, \nu=1/2, \eta=0 $ universality
   class, while the second is 1st order.
   The nature of the transition from the VB-SS to the SF where $ < \psi_{a} > = < \psi_{b} >
   =0 $ and  $ < \psi^{\dagger}_{a} \psi_{b} > = 0 $  inside the window driven by
   the quantum fluctuation $ r $ in the Fig.2b will be studied in
   \cite{un}.


{\sl 6.  Square lattice. }
     As said below Eqn.\ref{away}, with correspondingly defined $ \psi_{a/b} $ in a square lattice,
     upto the quartic level, Eqn.\ref{away} is the same as that in the
     square lattice derived in \cite{pq1}. So in the Ising limit,
     Fig.2b remains the same as Fig.1b.
     However, in the Easy-plane limit, as shown in \cite{pq1}, the lowest order
     term coupling the two phases $ \theta_{a/b} $ is
     $  C_{sq}= \lambda \cos 4 \theta $. If $ \lambda $ is positive ( negative),
     the VBS is Columnar dimer ( plaquette ) pattern. So the $ C_{3} $ term in
     Eqn.\ref{ep1} need to be replaced by $ C_{sq} $.
     Because $ C_{sq} $ is irrelevant near the QCP, the transition from the SF to the VBS
     could be a 2nd order transition through the deconfined QCP \cite{senthil} as shown in Fig.2c.
     We expect that the SF to
     the VB-SS transition could also be 2nd order through a novel deconfined quantum  {\em critical
     line } shown in Fig.2c and will be studied in \cite{un}.

\begin{figure}
\includegraphics[width=8cm]{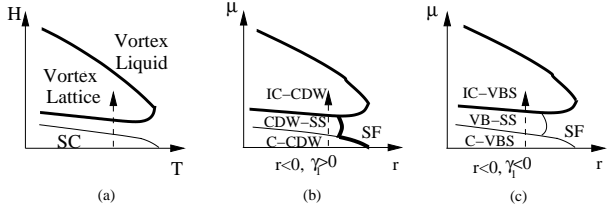}
\caption{(a) The phase diagram {\em slightly away}
    from $ q=1 $ is the same as type-II superconductor in external magnetic field $ H $  where there
    should be a vortex liquid state intruding between the Messiner state and the
    vortex lattice state. But the intruding regime is too narrow to be seen in type-II
    superconductors \cite{huse}. (b) and (c) are the  zero temperature phase diagrams of
    the chemical potential $ \mu $ versus $ r $ in Eqn.\ref{away} in square lattice.
    (b) The Ising limit $ \gamma_{1} > 0 $ is similar to Fig.1b.
    (c) The Easy-Plane limit $ \gamma_{1} < 0 $ is similar to Fig.1c
    except the SF to the C-VBS and the SF to the VB-SS  could be 2nd
    order  due to the so called "deconfined quantum critical point".
    (b) and (c) are drawn only above half filling. $ \mu \rightarrow - \mu $ corresponds to
    below half filling. The IC-CDW in (b) can be stabilized only by very long range
    interactions in Eqn.\ref{boson}. If it is not stable, then Fig.2b reduces to Fig.3b.
    The first order transition in (b) is a strong one.  }
\label{fig2}
\end{figure}

 {\sl 7. Implication on Quantum Monte-Carlo (QMC) simulations.}
   QMC simulations of hard core bosons on square lattice with  $ V_{1} $ and
   $ V_{2} $ interactions find a stable striped $ ( \pi,0) $ and $ (0,\pi) $
   SS ( Fig.3b ) \cite{hard}. A stable $ ( \pi, \pi) $ SS \cite{soft} can be realized
   in soft core boson case.
   But the nature of the CDW to supersolid transition has never
   been addressed. Our results show that the CDW to the SS transition must be in
   the same universality class of Mott to superfluid transition with exact exponents $ z=2, \nu=1/2, \eta=0
   $ with logarithmic corrections. It is important to  (1) confirm this prediction by finite
   size scaling through the QMC simulations in square lattice for $ (0,\pi) $ and $ (\pi,0 ) $
   supersolid in hard core case (Fig.3b) and $ (\pi,\pi) $ supersolid in the soft core case
   (2) do similar things in honeycomb lattice to confirm Fig.1.
   (3) To Eqb.\ref{boson} with $ U=\infty, V_{1} > 0 $, adding ring exchange term $ -K_{s} \sum_{ijkl} ( b^{\dagger}_{i}
   b_{j} b^{\dagger}_{k} b_{l} + h.c. ) $ \cite{sand}  where $ i, j, k, l $ label
   4 corners of a square in the square lattice and $ -K_{h} \sum_{ijklmn} ( b^{\dagger}_{i}
   b_{j} b^{\dagger}_{k} b_{l} b^{\dagger}_{m} b_{n} + h.c. ) $ where $ i, j, k, l, m, n $ label
   6 corners of a hexagon in the honeycomb lattice to stabilize the C-VBS state {\em  at half filling
   } ( $ K_{s}, K_{h} > 0 $ are free of sign probelm in QMC ),
   then confirm the prediction on C-VBS to VB-SS transition in Fig.2c and Fig.1c.
   The second transition ( CDW-SS to IC-CDW in Fig. 1b or 2b and the VB-SS to
   IC-VBS in Fig 1c or
   Fig.2c ) is hard to be tested in QMC, because some very long range
   interactions are needed to stabilize the IC-CDW or the IC-VBS
   state. They are first order transition anyway.

\begin{figure}
\includegraphics[width=8cm]{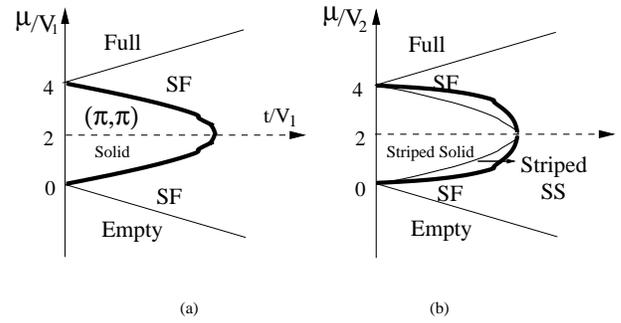}
\caption{( The thin (thick) line
    is a 2nd (1st ) order transition.   (a) Phase diagram for $ V_{2}=0,
    U=\infty $, the $ (\pi,\pi) $ SS, having negative compressibility,
    is unstable against phase separation
    into the SF and solid \cite{add}. The DVM does not apply in this case.
    (b) Phase diagram for $ V_{1}=0,U=\infty $, there is a narrow window of $ (
    \pi,0 ) $ SS sandwiched between the $ ( \pi,0 ) $ solid and
    the SF. The SS disappears at half-filling.
    The DVM does apply in this case. It corresponds to Fig.2b from the DVM. }
\label{fig3}
\end{figure}

   In fact, one of the predictions in this letter on the scaling of the superfluid
   density $ \rho_{s} \sim | \rho-1/2| $  was
   already found in the striped $ (\pi,0) $ solid to striped
   supersolid transition by QMC in Sec.V-B \cite{add}. In fact, as shown in section 4, there should be
   logarithmic correction to the scaling of $ \rho_{s} $, it
   remains a challenge to detect the logarithmic correction in QMC.
   Of course, the superfluid density is anisotropic $ \rho^{x}_{s}> \rho^{y}_{s} $ in the $ (\pi,0) $
   solid, but they scale in the same way with different coefficients
   \cite{add}. Although the
   authors in \cite{add} suggested it is a 2nd order transition, they did not
   address the universality class of the transition.

{\sl 8. Summary } We studied some phases and phase transitions in an
extended boson Hubbard model
  near half filling on bipartite lattices such as honeycomb and square lattice.
   We identified boson density and boson kinetic energy operators
   to characterize symmetry breaking patterns in the insulating states and supersolid states.
   We found that in the insulating side, the transition at zero temperature  driven by the chemical potential must be
   a C-CDW ( or C-VBS ) at half filling
   to a narrow window of CBW- ( VB-) supersolid,
   then to a IC-CDW ( IC-VBS ) transition in the Ising ( easy-plane ) limit.
   The valence bond supersolid is a new kind of supersolid first proposed in this letter.
   The first transition is in the  same universality class as that from a Mott insulator to
   a superfluid driven by a chemical potential, therefore have exact exponents $ z=2,
   \nu=1/2, \eta=0 $ with logarithmic corrections. The second is a 1st order transition.
   The results achieved in this letter could guide QMC
   simulations to search for all these phases and confirm the
   universality class of the transitions.
   These transitions could be easily realized in near future atomic experiments in optical lattices.


\begin{thebibliography}{99}
\bibitem{equal} The discussions  in the following equally apply to the striped solid with $
(\pi,0) $ or $ (0,\pi) $ order with the exception of the anisotropy
in the superfluid density. The anisotropy is irrelevant to the
universality class.
\bibitem{direct} In the direct lattice picture, we can simply renormalize away the
sublattice  $ A $ where the bosons sit and focus on the effective
boson hopping on the sublattice $ B $ whose dual vortex action is
given by Eqn.\ref{is}.
\end{thebibliography}
\end{document}